\documentclass[aps,prl,twocolumn,groupedaddress,superscriptaddress,amsfonts,amssymb,amsmath,citeautoscript,a4paper]{revtex4-1}

\usepackage[utf8]{inputenc}
\usepackage[english]{babel}
\usepackage{microtype}
\usepackage{txfonts}
\usepackage{txfontsb}

\usepackage{bm}
\usepackage{xcolor}
\usepackage{graphicx}
\usepackage{float}
\usepackage{xspace}
\usepackage{etoolbox}

\usepackage{enumerate}
\usepackage[inline]{enumitem}

\usepackage{siunitx}
\usepackage[pdftex]{hyperref}
\hypersetup{colorlinks,
            linkcolor={blue!75!black!80!yellow},
            citecolor={blue!75!black!80!yellow},
            urlcolor={blue!75!black!80!yellow}}

\usepackage[eulergreek]{sansmath} 
\makeatletter
\renewcommand\@make@capt@title[2]{%
    \@ifx@empty\float@link{\@firstofone}{\expandafter\href\expandafter{\float@link}}%
    \sisetup{math-sf=\textsf}%
    \sansmath\sffamily\textbf{#1\@caption@fignum@sep}#2
}%

\makeatother

\newcommand{\e}{\mathrm{e}}
\newcommand{\iu}{\mathrm{i}}
\newcommand{\ep}{\epsilon}
\newcommand{\epsm}{\epsilon_\mathrm{m}}
\newcommand{\epsb}{\epsilon_\mathrm{b}}
\newcommand{\epsd}{\epsilon_\mathrm{d}}
\newcommand{\kzd}{k_{z,\mathrm{d}}}

\newcommand{\GammaEELS}{\Gamma_\text{\textsc{EELS}}}
\newcommand{\GammaCL}{\Gamma_\text{\textsc{CL}}}
\renewcommand{\Re}{\operatorname{Re}}
\renewcommand{\Im}{\operatorname{Im}}
\newcommand{\appropto}{\mathrel{\vcenter{
			\offinterlineskip\halign{\hfil$##$\cr
				\propto\cr\noalign{\kern.5pt}\sim\cr\noalign{\kern-1pt}}}}}

\newcommand{\ICFOaffil}{ICFO -- Institut de Ci{\`e}ncies Fot{\`o}niques, The Barcelona Institute of Science and Technology, 08860 Castelldefels, Barcelona, Spain}
\newcommand{\ICREAaffil}{ICREA -- Instituci\'o Catalana de Recerca i Estudis Avan\c{c}ats, Passeig Llu\'{\i}s Companys 23, 08010 Barcelona, Spain}

\begin{document}
\title{Interrogating Quantum Nonlocal Effects in Nanoplasmonics through Electron-Beam Spectroscopy}
%
\author{P.~A.~D.~Gon\c{c}alves}
\email{andre.goncalves@icfo.eu}
\affiliation{\ICFOaffil}
\author{F.~Javier~Garc\'{\i}a~de~Abajo}
\email{javier.garciadeabajo@nanophotonics.es}
\affiliation{\ICFOaffil}\affiliation{\ICREAaffil}


\begin{abstract}
 A rigorous account of quantum nonlocal effects is paramount for understanding the optical response of metal nanostructures and for designing plasmonic devices at the nanoscale. Here, we present a scheme for retrieving the quantum surface response of metals, encapsulated in the Feibelman $d$-parameters, from electron energy-loss spectroscopy (EELS) and cathodoluminescence (CL) measurements. We theoretically demonstrate that quantum nonlocal effects have a dramatic impact on EELS and CL spectra, in the guise of spectral shifts and nonlocal damping, when either the system size or the inverse wave vector in extended structures approach the nanometer scale. Our concept capitalizes on the unparalleled ability of free-electrons to supply deeply subwavelength near-fields and, thus, probe the optical response of metals at length scales in which quantum-mechanical effects are apparent. These results pave the way for a widespread use of the $d$-parameter formalism, thereby facilitating a rigorous yet practical inclusion of nonclassical effects in nanoplasmonics.
\end{abstract}

\maketitle


The optical response of few-nanometer-scale plasmonic structures, such as those engineered with state-of-the-art nanofabrication techniques, can exhibit substantial quantum nonlocal effects associated with the inherently quantum mechanical nature of the plasmon-supporting electron gas in the involved materials\cite{ZEB16,F1982,KV95,CHM12,RTP12,SKD12,RSK13,RKC15,TNA13,VGF16,ZFR14,CYJ17,CTC19,YZY19,GCR20,BDC21}. 
Broadly speaking, the impact of nonclassical effects becomes non-negligible when either the characteristic size of the system falls below $\SIrange{\sim 10}{20}{\nm}$ or the optical response is mediated by field components of large momenta such as those produced by confined near-field confinement. Hence, a quantum nonlocal description of the underlying plasmon-mediated light--matter interaction is required in order to explain experimental data as well as to draw insight into the elementary processes governing that interaction in the few-nanometer regime. 

Since an all-encompassing quantum-mechanical treatment of the many-electron system [e.g., using  time-dependent density-functional theory~\cite{MUN06} (TDDFT)] is severely constrained to few-atom clusters much smaller than the typical nanoplasmonic systems exploited in experiments, in practice it is necessary to resort to quantum-informed models that incorporate dominant quantum effects to leading-order~\cite{ZEB16,PSC07,paper178,M21}. 
Among these, the Feibelman $d$-parameter formalism~\cite{F1982} is particularly appealing because it simultaneously incorporates electron spill-out/spill-in, nonlocality (i.e., momentum-dependent response), and surface-enabled Landau damping through the introduction of two microscopic surface-response functions, $d_\perp(\omega) = \int \mathrm{d}z\, z\, \rho_\mathrm{ind}(z,\omega) / \int \mathrm{d}z\, \rho_\mathrm{ind}(z,\omega)$ and $d_\parallel(\omega) = \int \mathrm{d}z\, z\, \partial_z J_{\parallel,\mathrm{ind}}(z,\omega) / \int \mathrm{d}z\, \partial_z J_{\parallel,\mathrm{ind}}(z,\omega)$, corresponding to the centroids of the induced charge density along the surface normal $\mathbf{\bf\hat{z}}$ and of the normal derivative of the current parallel to the interface, respectively. Once they are known for the planar dielectric--metal interface(s) of interest, these parameters allow the incorporation of the above-mentioned nonclassical effects in the optical response of metallic nanostructures using standard electromagnetic solvers upon replacing the macroscopic boundary conditions~\cite{J1975} by their $d$-parameter-corrected counterparts~\cite{YZY19,GCR20,paper375,ZZC22,YQ22,HU22,BOS22,TZL22}. Naturally, this procedure relies on our ability to compute the $d$-parameters in the first place using, for example, linear-response TDDFT. 
However, while simple metals (e.g., alkali metals or aluminum) can be well-described by jellium-level TDDFT, for which accurate $d$-parameter data exist~\cite{F1982,L1987,KLS1988,L97_3,CYJ17}, noble metals such as gold and silver require a more demanding atomistic treatment beyond the jellium approximation due to valence-electron screening from the lower-lying bands~\cite{L97_3,STP1989}. 
While valence-band screening may be semiclassically included using screened jellium models~\cite{L93,L97_3} containing a polarizable background contribution, such approaches still lead to \emph{quantitatively} unsatisfactory predictions for the $d$-parameters (see, for instance, the discussion in the supplementary information of refs.~\citenum{CYJ17} and \citenum{YZY19}). 
As a result of this, and despite the relevance of noble metals in nanoplasmonics, quantitatively accurate $d$-parameter data remains elusive, thus limiting the widespread use of the $d$-parameter framework.

\begin{figure*}
 \centering
  \includegraphics[width=0.8\textwidth]{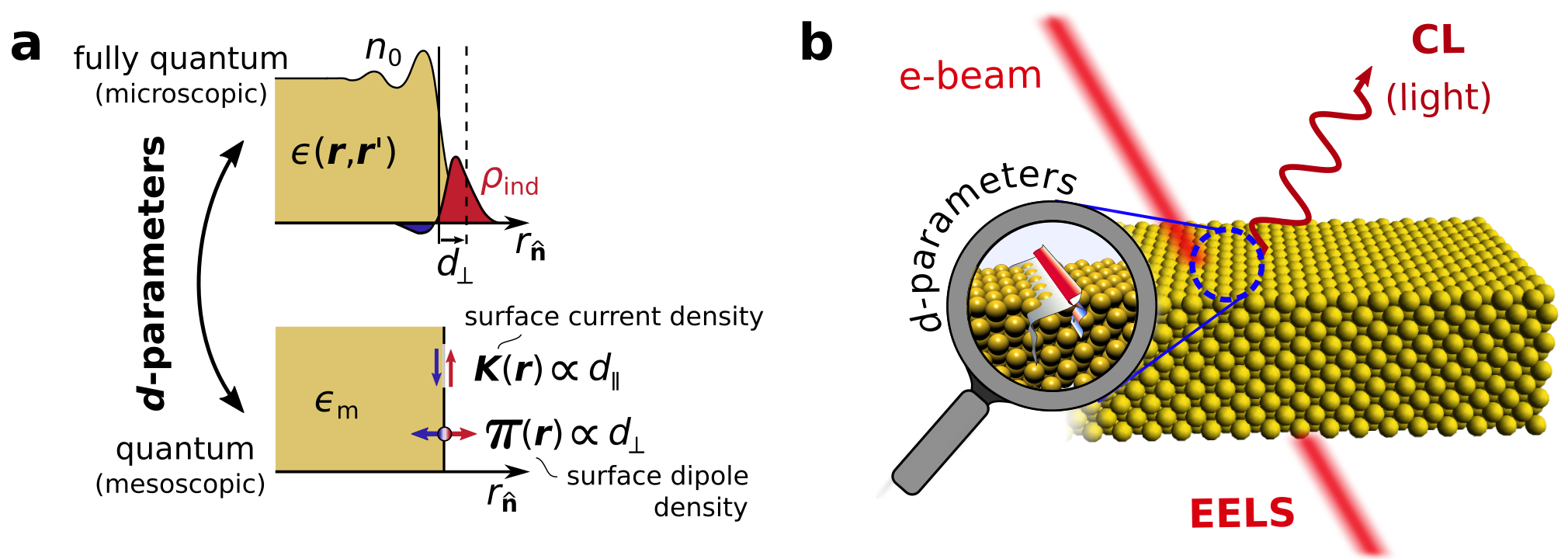}
  \caption{\textbf{Probing quantum effects in nanoplasmonics with electron-beam (e-beam) spectroscopy.} 
  \textbf{(a)}~Conceptual approach underpinning the Feibelman $d$-parameter formalism, wherein a microscopic, quantum mechanical description of a dielectric--metal interface is mapped onto a mesoscopic one that is tantamount to a classical treatment augmented by a set of quantum surface-response functions, $d_\perp$ and $d_\parallel$, encapsulating the leading-order corrections to classicality.
  \textbf{(b)}~Schematics of metallic quantum surface response encoded in the $d$-parameters and probed via EELS and CL spectroscopies.
  }
  \label{fig:1}
\end{figure*}

Here, we propose and demonstrate a scheme in which electron-beam (e-beam) spectroscopies~\cite{paper149,paper338} are employed to determine the quantum surface response (i.e., the $d$-parameters) of metals directly from experimental spectra (Fig.~\ref{fig:1}). To that end, we present a quantum-corrected theory of electron energy-loss spectroscopy~\cite{E96,paper149,paper338} (EELS) and cathodoluminescence~\cite{KZ17,paper149,paper338} (CL) based on the aforementioned quantum surface-response formalism and use it to infer $d_\perp$ and $d_\parallel$ from the measured spectra by quantifying the size- or wave-vector-dependent spectral shifting and broadening due to quantum nonlocal effects. Crucial to this is the ability of e-beams to produce broadband and highly confined near-fields~\cite{paper149}, which may be tailored by, for example, varying the electron kinetic energy or  controlling the e-beam trajectory. 
Such fields contain evanescent components that allow free electrons to efficiently couple to strongly confined optical excitations in materials and retrieve sub-nanometer spatial information, thus rendering them first-class probes of nonclassical effects in nanoplasmonics~\cite{SKD12,RSK13,RKC15,CTC19}. 
Our work opens an powerful route toward a better quantitative understanding of the nonclassical optical response of metallic nanostructures, which is instrumental from a fundamental viewpoint and constitutes a key ingredient in the design of nanophotonic devices operating at the few-nanometer scale.


We begin our analysis by considering the canonical scenario of a swift electron moving with constant velocity $v$ along a straight-line trajectory $\mathbf{r}_e(t)$ parallel to a metal surface placed at $z=0$. 
Taking $\mathbf{v} = v\, \mathbf{\hat{x}}$ and $\mathbf{r}_e(t) = (v\, t,0,b)$, with $b$  defining the electron--surface separation, and assuming that the medium adjacent to the metal is a lossless dielectric with relative permittivity $\epsd$, the spectral EELS probability experienced by the electron after traveling a length $L$ reads~\cite{paper149}
\begin{align}
 \GammaEELS (\omega) = \frac{2 \alpha L}{\pi c\, \beta^2} \int_{0}^{\infty} \frac{\text{d}k_y}{q^2} 
 \Re \left\{ \e^{\iu 2 \kzd b} \left[  
 \frac{k_y^2 \, \beta^2}{\kzd} r_s - \frac{\kzd}{\epsd} r_p 
 \right] \right\} 
 , \label{eq:EELS_planar}
\end{align}
where $\beta = v/c$ is the normalized electron velocity, $\alpha \simeq 1/137$ the fine-structure constant, and $q = \sqrt{\omega^2/v^2 + k_y^2}$ and $\kzd = \sqrt{\epsd\, \omega^2/c^2 - q^2}$ (with $\Im \{\kzd\} \geq 0$) stand for the in-plane and out-of-plane wave vector components, respectively. The quantum surface response enters Eq.~\eqref{eq:EELS_planar} through the reflection coefficients for $s$- and $p$-polarized waves, $r_s \equiv r_s(q,\omega)$ and $r_p \equiv r_p(q,\omega)$, respectively. The EELS probability expressed in the form of Eq.~\eqref{eq:EELS_planar} is thus particularly convenient to incorporate quantum nonlocal effects by simply employing the $d$-parameter-corrected reflection coefficients~\cite{F1982,YZY19,GCR20,G20} (see Methods) instead of their classical counterparts, which are reinstated in the $d_{\perp,\parallel} \to 0$ limit.

Incidentally, $d_\parallel$ vanishes for charge-neutral surfaces~\cite{F1982,L97_3}, thereby leaving $d_\perp$ as the only quantity embodying quantum mechanical corrections in the present context, where we take $d_\parallel = 0$. We consider both jellium-like and noble metals (as their nonclassical optical response is distinct), herein represented, respectively, by a jellium with density parameter $r_s=4$ (corresponding to the plasma energy $\hbar\omega_{\mathrm{p}} \approx \SI{5.89}{\eV}$ for sodium~\cite{AM1976}) and silver. 
For the former, we use the frequency-dependent $d_\perp$ calculated from TDDFT~\cite{CYJ17} for an air--jellium interface (see SI), whereas for silver we incorporate a surrounding dielectric with $\epsd=2$ (simulating $\mathrm{SiO}_2$, which protects it from oxidation) and take $d_\perp = (-0.4 + 0.2\, \iu) \, \mathrm{nm}$. This value is estimated by fitting its real part to experimental measurements of size-dependent resonance shifts~\cite{RSK13}, while its imaginary part is set so that it reproduces the phenomenological Kreibig damping~\cite{KF1969} (see SI for details). 
The classical optical response of silver is modeled through a Drude-type dielectric function $\epsm(\omega) = \epsb(\omega) - \omega_{\mathrm{p}}^2/(\omega^2 + \iu \gamma \omega)$, where $\hbar\omega_{\mathrm{p}} = \SI{9.02}{\eV}$ and $\hbar\gamma = \SI{22}{\meV}$ describe the conduction electrons, whereas screening due to bound electrons is included via $\epsb(\omega) = \epsm^{\mathrm{exp}}(\omega) + \omega_{\mathrm{p}}^2/(\omega^2 + \iu \gamma \omega)$ with $\epsm^{\mathrm{exp}}(\omega)$ taken from experimental data~\cite{JC1972}.

\begin{figure*}
 \centering
  \includegraphics[width=1.0\textwidth]{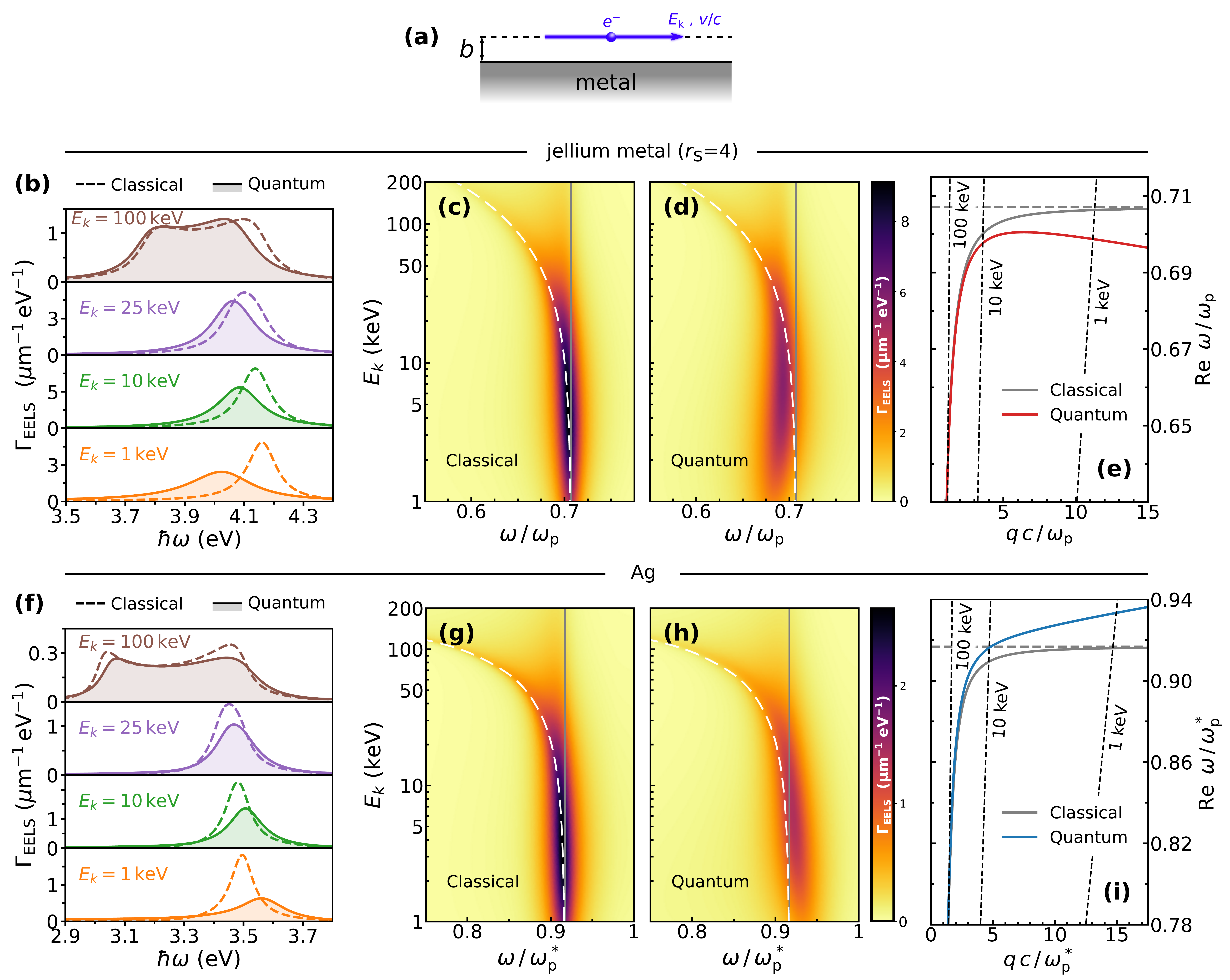}
  \caption{\textbf{Nonclassical corrections to the EELS for an aloof electron parallel to a planar metal surface.} 
  \textbf{(a)}~Schematics of the configuration under consideration. 
  \textbf{(b)}~Classical and quantum EELS spectra for an electron traveling in air ($\epsd=1$) above a planar jellium surface ($r_s=4$, with $\hbar\omega_{\mathrm{p}} \simeq \SI{5.89}{\eV}$ and $\gamma=\omega_{\mathrm{p}}/50$) for selected values of the kinetic energy $E_k$.
  \textbf{(c,d)}~Classical (c) and quantum (d) EELS spectra for the same air--jellium interface as a function of $E_k$. The classical SPP result at $q=\omega/v$ is indicated by the white-dashed curves, while the vertical gray-solid lines indicate the classical nonretarded surface plasmon frequency $\omega_{\mathrm{SP}}^{\mathrm{cl}} = \omega_{\mathrm{p}}/\sqrt{2}$. 
  \textbf{(e)}~Dispersion relation of SPPs from classical and quantum treatments of the planar air--jellium interface in (b--d). 
  \textbf{(f--i)}~Same as (b--e), but for a silver surface (screened plasma frequency $\hbar\omega_{\mathrm{p}}^{*} = \SI{3.82}{\eV}$) capped with a dielectric of permittivity $\epsd=2$ (representative of $\mathrm{SiO}_2$). We take $b=\SI{5}{\nm}$ in all cases.
  }
  \label{fig:2}
\end{figure*}

The impact of quantum nonlocal effects imparted on the EELS spectrum of an electron traveling parallel to a planar metal surface is presented in Fig.~\ref{fig:2} (see panel (a) for a sketch of the geometry). Notably, while at large electron kinetic energies $E_k$ the EELS spectra are well-described by classical dielectric theory, such a description progressively deteriorates as $E_k$ is reduced. 
More precisely, we find that for $E_k \lesssim \SI{20}{\keV}$ the impact of nonclassical effects becomes substantial, imprinting considerable spectral shifts and resonance broadening on the EELS spectra. The broadening is a direct consequence of surface-assisted Landau damping, entering via $\Im \{d_\perp\}$, whereas the observed resonance shifts are produced by the displacement of the induced charges relative to the classically defined abrupt interface, which is encoded by $\Re \{d_\perp\}$. The sign of $\Re \{d_\perp\}$ dictates the direction of the frequency shift: toward the red if positive, reflecting the electron spill-out characteristic of jellium metals (Fig.~\ref{fig:2}b--e)~\cite{L1987,TPF1989,TPL91,SWP92,RES95}; or toward the blue if negative, signaling the electron spill-in observed in silver (Fig.~\ref{fig:2}f--i) and other noble metals~\cite{L93,TKM93,CCL06,CHM12,SKD12,RSK13,RKC15,YZY19,CTC19}.
Furthermore, since the peak in the EELS spectrum is associated with the excitation of surface plasmon polaritons (SPPs), the observation that the impact of quantum nonlocal effects grows with decreasing $E_k$ can be understood as follows: (i) the main contribution to the EELS probability arises at a lost energy $\hbar\omega$ for which the wave-vector transfer threshold $q=\omega/v$ intersects that of the SPP; (ii) lower electron velocities lead to intersections occurring at correspondingly larger wave vectors (Figs.~\ref{fig:2}e and~\ref{fig:2}i), which is precisely where quantum nonlocal effects become sizable (with resonance frequency shifts $\appropto q \Re\{d_\perp\}$ and nonlocal broadening $\appropto q \Im\{d_\perp\}$)~\cite{CYJ17,GCR20,G20}. Together, (i) and (ii) provide a simple and intuitive explanation underpinning the main features observed in Fig.~\ref{fig:2}.

\begin{figure*}
 \centering
  \includegraphics[width=1.0\textwidth]{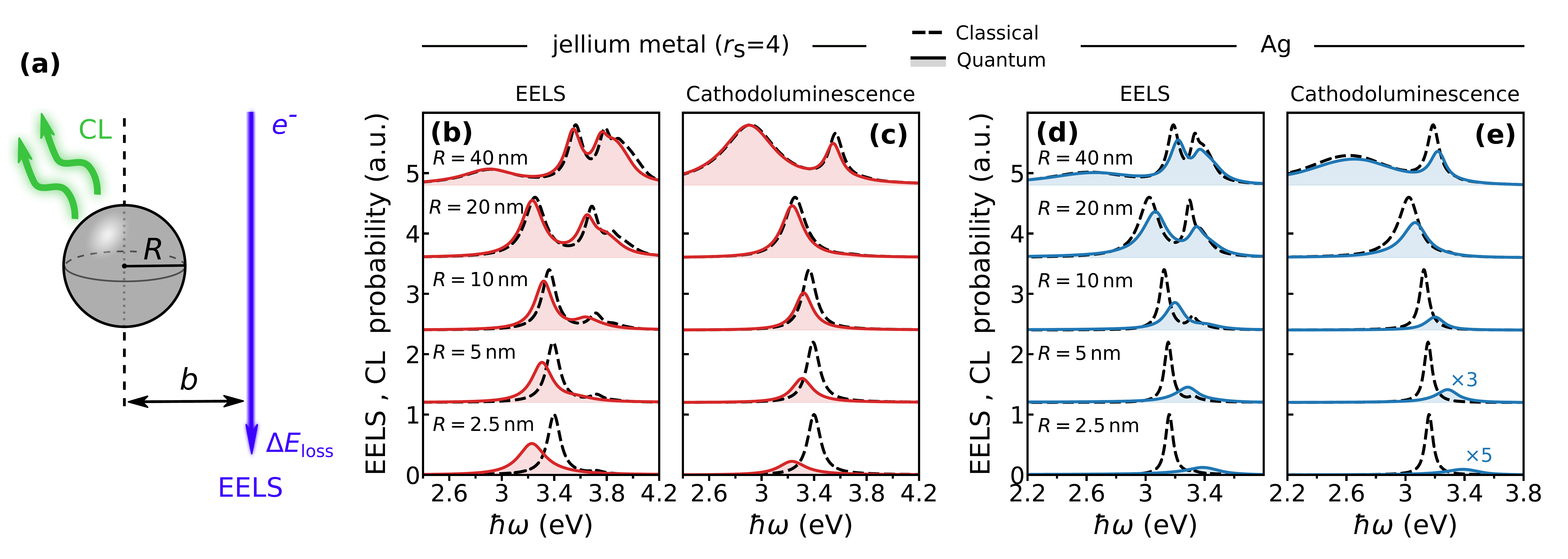}
  \caption{\textbf{Nonclassical optical response of metallic spheres probed through EELS and CL spectroscopies.} 
  \textbf{(a)}~Illustration of the aloof configuration under consideration. 
  \textbf{(b,c)}~Calculated EELS (b) and CL (c) spectra for jellium spheres with different radii in air, contrasting the classical (black dashed curves) and quantum (color-filled solid curves) treatments. 
  \textbf{(d,e)}~Same as (b,c), but for silver spheres in a host dielectric with $\epsd=2$.  We take $E_k = \SI{50}{\keV}$ and $b=R+\SI{5}{\nm}$ in all cases.
  }
  \label{fig:3}
\end{figure*}

Metal nanoparticles constitute another quintessential architecture in which e-beam spectroscopies have played an important role (e.g., to map plasmonic fields in real-space with nanometric resolution~\cite{paper085,KS14,KZ17}). As we show below, localized surface plasmon (LSP) resonances in small metal nanoparticles investigated with EELS and/or CL can be used to quantitatively probe the nonclassical optical response of metals. Focusing on metal spheres, the spectrally resolved EELS and CL probabilities associated with an aloof e-beam passing near a sphere of radius $R$ with impact parameter $b>R$ (see Fig.~\ref{fig:3}a) are given by
\begin{align}
 \GammaEELS (\omega) = \frac{\alpha}{\omega \sqrt{\epsd}} 
 &\sum_{l=1}^{\infty} \sum_{m=-l}^{l} K_m^2 \left( \frac{\omega b}{v\, \gamma_{\epsd}} \right) \nonumber\\[0.25em]
 \times &\left[
 C_{lm}^{\text{\textsc{E}}}(\beta_{\epsd})\, \Im \left\{ t_l^{\text{\textsc{E}}} \right\}
 + C_{lm}^{\text{\textsc{M}}}(\beta_{\epsd})\, \Im \left\{ t_l^{\text{\textsc{M}}} \right\}
 \right] 
 , \label{eq:EELS_sphere}
\end{align}
and 
\begin{align}
 \GammaCL (\omega) = \frac{\alpha}{\omega \sqrt{\epsd}} 
 &\sum_{l=1}^{\infty} \sum_{m=-l}^{l} K_m^2 \left( \frac{\omega b}{v\, \gamma_{\epsd}} \right) \nonumber\\[0.25em] 
 \times &\left[
 C_{lm}^{\text{\textsc{E}}}(\beta_{\epsd})\, \left| t_l^{\text{\textsc{E}}} \right|^2 
 + C_{lm}^{\text{\textsc{M}}}(\beta_{\epsd})\, \left| t_l^{\text{\textsc{M}}} \right|^2
 \right] 
 , \label{eq:CL_sphere}
\end{align}
respectively, where $K_m$ is a modified Bessel function of the second kind~\cite{AS1972}, $\gamma_{\epsd} = (1 - \beta_{\epsd}^2)^{-1/2}$, and we have defined $\beta_{\epsd} = \sqrt{\epsd} v/c$. Here, the quantities $C_{lm}^{\text{\textsc{E}}}$ and $C_{lm}^{\text{\textsc{M}}}$ are coupling coefficients that, for a given pair of angular momentum numbers $(l,m)$, depend uniquely on $\beta_{\epsd}$ (see Ref.~\cite{paper149} for explicit expressions). 
Equations~\eqref{eq:EELS_sphere} and \eqref{eq:CL_sphere} extend the previously derived results for the interaction of a fast electron with a sphere in vacuum~\cite{paper021,paper149} to a configuration in which the sphere is embedded in a lossless dielectric medium with arbitrary $\epsd$. The optical response of the sphere enters these equations through the Mie scattering coefficients $t_l^{\text{\textsc{E}}}$ and $t_l^{\text{\textsc{M}}}$ for transverse magnetic (TM) and transverse electric (TE) waves, respectively. 
In analogy to the planar interface considered above, quantum mechanical corrections in the optical response are straightforwardly accounted for by adopting the generalized Mie coefficients containing the $d$-parameters~\cite{GCR20} (see Methods).

Figure~\ref{fig:3} compares classical and quantum predictions for the EELS probability (Figs.~\ref{fig:2}b,d) and CL (Figs.~\ref{fig:2}c,e) spectra from metallic spheres with different radii. In many ways, they echo the general conclusions discussed above for the planar interface, but in this instance $R^{-1}$ takes the role previously played by the in-plane wave vector $q$. Specifically, the nonclassical spectral shifts and broadening increase when reducing the particle radius---qualitatively following $\appropto l(l+1)\Re \{d_\perp\}/R$ and $\appropto l(l+1)\Im \{d_\perp\}/R$, respectively~\cite{GCR20}---, ultimately leading to pronounced differences in the spectral peak corresponding to the dipolar ($l=1$) LSP for $R \lesssim \SI{10}{\nm}$. 
In passing, we note that higher-order multipoles in larger spheres can still display deviations from classicality (profiting from the $l(l+1)$ factor noted above, which reflects the faster surface oscillations as $l$ increases), albeit much less recognizable in comparison to those observed for the dipolar LSP in small spheres. 
Indeed, aside from being quenched by nonlocal broadening, dipolar LSP resonances in jellium (silver) spheres of a few nanometers in size are dramatically red (blue) shifted (by as much as $\SI{\sim 200}{\meV}$) with respect to the classical nonretarded result $\omega^{\text{cl}} = \omega_{\text{p}}/\sqrt{\epsb + 2 \epsd}$. 
The breaking of scale-invariance characterizing the classical nonretarded limit is thus lifted within this investigated regime due to the introduction of the inherently quantum-mechanical length-scale associated with $|d_\perp|$. 

Although nonclassical effects permeate EELS and CL spectra in similar ways, there are some important differences. Being the result of spontaneous light emission following e-beam excitation, CL is only sensitive to bright LSP modes, whereas EELS grants us access to dark multipolar LSPs~\cite{LK15,paper251} (cf. the EELS and CL spectra in Fig.~\ref{fig:3}). In addition, the CL signal drops considerably for small nanoparticles due to the realization of the dipole limit and the concomitantly smaller scattering cross section. Therefore, EELS is better suited for measuring the optical response at very small sizes, with EELS measurements of silver particles down to $\SI{\sim 2}{\nm}$ in diameter having been reported~\cite{SKD12,RKC15,CTC19}. 

\begin{figure*}
 \centering
  \includegraphics[width=0.8\textwidth]{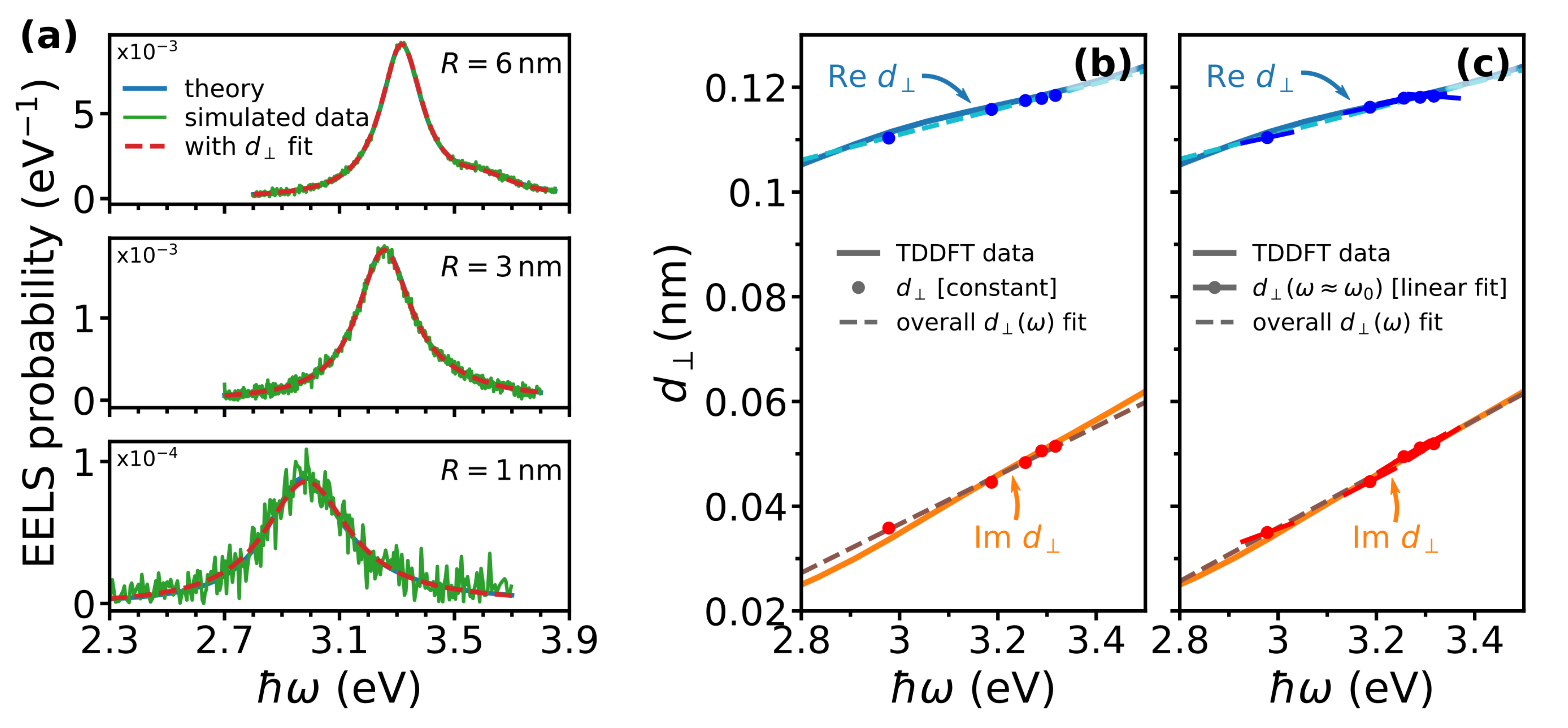}
  \caption{\textbf{Retrieval of $\bm d_\perp(\bm\omega)$ from (simulated) experimental EELS data of metallic jellium spheres in air.} 
  \textbf{(a)}~EELS spectra for selected radii (indicated in each panel), showing the theoretically calculated result (corresponding to the ``ground truth'', blue curves), the simulated experimental data (by adding noise to the previously calculated ones, green curves), and the reconstructed spectra by fitting $d_\perp$ (dashed red curves).
  \textbf{(b--c)}~Extracted $d_\perp$-parameter data from EELS spectra of jellium nanospheres of different radii. Each dot (one for the real part and another one for the imaginary part) corresponds to the fitted value of $d_\perp$ for a specific radius $R=1,2,3,4,6\, \mathrm{nm}$ (each corresponding to a specific LSP resonance frequency $\omega_0$). The dashed lines are obtained from a linear fit for $d_\perp(\omega)$ using all the data for different radii. 
  In (b), for each radius, $d_\perp$ is fitted to a constant, whereas in (c) it is fitted to a linear frequency dependence around each $\omega_0$. 
  The solid lines represent the original TDDFT data (same as in Fig.~S1) in the spectral window of interest for LSPs in jellium sheres. 
  In all cases, we take $b=R+\SI{5}{\nm}$ and $E_k = \SI{50}{\keV}$.
  }\label{fig:4}
\end{figure*}

Furthermore, focusing on metallic jellium spheres for concreteness, we explicitly show in Fig.~\ref{fig:4} a scheme to extract $d$-parameter data from EELS measurements (for CL the procedure and results would be the same) using the $d$-parameter-corrected theory introduced above (further details on the implementation are provided in Supplementary Section 2). 
We emulate experimental EELS data by adding noise to the theoretically calculated spectra (Fig.~\ref{fig:4}a, green curves), while also accounting for the fact that the EELS signals from smaller nanoparticles exhibit more noise than those from larger ones due to the smaller signal-to-noise ratio. We stress that the simulated experimental spectra mimics well those obtained from actual EELS measurements~\cite{CTC19}. We then use our $d$-parameter-corrected theory to extract $d_\perp$ following two alternative approaches: a simple one in which we fit a constant $d_\perp$ value for each particular radius (Fig.~\ref{fig:4}b), and a more refined one in which we infer the $d_\perp$ dispersion near the LSP peak to leading order via a linear fit (Fig.~\ref{fig:4}c). Finally, the combination of data from spheres of different radii enables the reconstruction of the complex-valued, frequency-dependent $d_\perp$ across a broad spectral range with extremely good accuracy (Fig.~\ref{fig:4}b--c, dashed lines), thereby underscoring the viability of our proposed scheme. 
Importantly, such a scheme is extremely valuable to unravel the quantum surface response of metals directly from experimental data with high accuracy. This is particularly relevant for noble metals, for which quantitatively accurate $d$-parameter calculations are not yet available. In addition, even for simple metals, for which TDDFT data exist, the determination of the $d$-parameters using the approach developed here provides a new path for benchmarking first-principles calculations (inasmuch as the fidelity of such methods ultimately depends on the chosen implementation or functional~\cite{LBB16,MS17}) directly against experiment. 
Incidentally, although we have employed a standard least-squares fitting procedure to extract the $d$-parameters (see SI), our approach could benefit from machine-learning methods, which have been applied in similar settings, namely, in transmission electron microscopy~\cite{SOJ21}, scanning near-field optical microscopy~\cite{CXS22}, ellipsometry~\cite{LZY21,WCZ22}, and others~\cite{PSJ18}.


In conclusion, we have demonstrated that EELS and CL spectroscopies constitute powerful tools to probe quantum-mechanical corrections in nanoplasmonics, which here we have calculated by augmenting the classical, local-response theory with the Feibelman $d$-parameters. In particular, we have shown that quantum effects in the response of metallic surfaces lead to substantial nonclassical shifts and nonlocal broadening of the EELS and CL spectral features associated with surface plasmon resonances. In extended planar metal surfaces, such deviations from classicality become non-negligible for electron kinetic energies below $\SI{\sim 20}{\keV}$ due to the contribution from large wave vector components associated with free-electrons, which increases as the electron energy is lowered. In metallic spheres, the relevant length scale is instead determined by the particle size, and thus, the impact of nonclassical corrections is weakly dependent on the electron kinetic energy (see Fig.~S3 in SI). Specifically, we find that quantum nonlocal effects become substantial for spheres with radii $\lesssim \SI{10}{\nm}$, in-line with experimental observations~\cite{SKD12,RKC15,CTC19}. 

Our work provides a viable, concrete scheme for interrogating the nonclassical optical response of metals in a quantitative fashion through the retrieval of the $d$-parameters associated with the involved dielectric--metal interfaces from EELS and CL measurements. In practice, as demonstrated here, this is achieved using the $d$-parameter-corrected theory to infer such parameters from fits of experimental spectra (Fig.~4), as all other experimental parameters can be well-characterized using currently available techniques. 
Additionally, we underscore that our proposal based on e-beam spectrocopies is superior to optics-based techniques because it can effectively address individual particles rather than an ensemble (which invariably introduces inhomogeneous broadening due to particle size and/or morphology variations). Yet another advantage of e-beams is that, by varying their orientation, tomography-based methods~\cite{NPL13,ABC15,HTH15,LHH21} can be exploited to precisely map the exact shape of the nanoparticle (and even account for surface roughness, a classical effect that could be erroneously mistaken for nonclassical spectral shifts and broadenings).

We envision that the scheme presented in this work will stimulate experimental endeavors for measuring the Feibelman $d$-parameters for relevant combinations of dielectric--metal interfaces. Indeed, a systematic compilation of a ``$d$-parameter catalogue'' would allow the full deployment of this formalism across the board in nanophotonics, with key implications not only for understanding the fundamentals of plasmon-based light--matter interactions at the nanoscale but also for optimizing and designing nanoplasmonic devices with nanometer-sized footprints.

\hfill

\section*{Methods}

{\small 
\noindent 
\textbf{Mesoscopic scattering coefficients.} %
The mesoscopic, $d$-parameter-corrected scattering coefficients for a planar metal surface and for metallic spheres have been previously introduced by Feibelman~\cite{F1982} and Gon\c{c}alves~et al.~\cite{GCR20}, respectively. Here, we reproduce them for completeness.

For the planar dielectric--metal interface, the nonclassical version of the Fresnel reflection coefficients for $p$- and $s$-polarized waves read~\cite{F1982,L97_3,GCR20,G20}
\begin{subequations}
\begin{align}
  r_p &= \frac{ \epsm k_{z,\mathrm{d}} - \epsd k_{z,\mathrm{m}} 
 + (\epsm - \epsd) \left[ \iu q^2 d_\perp - \iu k_{z,\mathrm{d}} k_{z,\mathrm{m}} d_\parallel \right] }{ \epsm k_{z,\mathrm{d}} + \epsd k_{z,\mathrm{m}} 
 - (\epsm - \epsd) \left[ \iu q^2 d_\perp + \iu k_{z,\mathrm{d}} k_{z,\mathrm{m}} d_\parallel \right] } , \\[0.25em]
 r_s  &=  \frac{k_{z,\mathrm{d}} - k_{z,\mathrm{m}} + (\epsm - \epsd) \iu k_0^2 d_\parallel}{k_{z,\mathrm{d}} + k_{z,\mathrm{m}} - (\epsm-\epsd) \iu k_0^2 d_\parallel} 
 ,
\end{align}%
\end{subequations}%
where $q$ is the in-plane wave vector, $k_0 = \omega/c$, and $k_{z,j} = \sqrt{\ep_j k_0^2 - q^2}$ with $j \in \{ \mathrm{m} , \mathrm{d} \}$ denoting the out-of-plane wave vector components.

For a metallic sphere of radius $R$, the generalized, nonclassical transverse magnetic (TM) and transverse electric (TE) Mie coefficients are given by~\cite{GCR20,G20}
\begin{widetext}
\begin{subequations}
\begin{align}
 t_l^{\text{\textsc{e}}} &= \iu\, \frac{%
 \epsm j_l(x_{\mathrm{m}}) \Psi'_l(x_{\mathrm{d}}) - \epsd j_l(x_{\mathrm{d}}) \Psi'_l(x_{\mathrm{m}}) 
 + 
 ( \ep_{\mathrm{m}} - \ep_{\mathrm{d}} ) \left\{ j_l(x_{\mathrm{d}}) j_l(x_{\mathrm{m}}) \left[l(l+1)\right] d_\perp + \Psi'_l(x_{\mathrm{d}}) \Psi'_l(x_{\mathrm{m}}) \, d_\parallel \right\} / R
 }{ %
  \epsm j_l(x_{\mathrm{m}}) \xi'_l(x_{\mathrm{d}}) - \epsd h_l^{(1)}(x_{\mathrm{d}}) \Psi'_l(x_{\mathrm{m}}) 
 + 
 ( \ep_{\mathrm{m}} - \ep_{\mathrm{d}} ) 
 \left\{ h_l^{(1)}(x_{\mathrm{d}}) j_l(x_{\mathrm{m}}) \left[l(l+1)\right] d_\perp + \xi'_l(x_{\mathrm{d}}) \Psi'_l(x_{\mathrm{m}}) \, d_\parallel \right\} / R
 } , \\[0.25em]
 t_l^{\text{\textsc{m}}} &= \iu\, \frac{%
	j_l(x_{\mathrm{m}}) \Psi'_l(x_{\mathrm{d}}) - j_l(x_{\mathrm{d}}) \Psi'_l(x_{\mathrm{m}})   
	+ \big( x_\mathrm{m}^2 - x_\mathrm{d}^2 \big) j_l(x_{\mathrm{d}}) j_l(x_{\mathrm{m}}) \, d_\parallel / R
}{%
	j_l(x_{\mathrm{m}}) \xi'_l(x_{\mathrm{d}}) - h_l^{(1)}(x_{\mathrm{d}}) \Psi'_l(x_{\mathrm{m}})  
	+ \big( x_\mathrm{m}^2 - x_\mathrm{d}^2 \big) \,h_l^{(1)}(x_{\mathrm{d}}) j_l(x_{\mathrm{m}}) \, d_\parallel / R
} ,
\end{align}%
\end{subequations}%
\end{widetext}
in terms of the dimensionless wave vectors $x_j \equiv  k_0 \sqrt{\ep_j} R$. Here, $j_l(x)$ and $h^{(1)}_l(x)$ stand for the spherical Bessel and Hankel functions of the first kind~\cite{AS1972}, $\Psi_l (x) \equiv x j_l (x)$ and $\xi_l (x) \equiv x h_l^{(1)} (x)$ are Riccati--Bessel functions~\cite{AS1972}, and primed functions denote their derivatives.
}


\section*{Supporting Information}

\noindent 
Supporting information is available free of charge at {DOI: \href{https://doi.org/10.1021/acs.nanolett.3c00298}{10.1021/acs.nanolett.3c00298}}. 
\begin{changemargin}{1em}{0em} 
Details on the Feibelman $d$-parameter data, comprehensive description of the scheme to extract the $d$-parameters from EELS spectra, and explicit demonstration of the robustness of the EELS and CL peak position with varying e-beam kinetic energy.
\end{changemargin}

\hfill
\section*{Acknowledgments}

\noindent 
This work has been partly supported by the European Research Council (Advanced Grant No.~789104-eNANO), the Spanish Ministry of Science and Innovation (PID2020-112625GB-I00 and CEX2019-000910-S), the Generalitat de Catalunya (CERCA and AGAUR), and the Fundaci\'{o}s Cellex and Mir-Puig. 


\providecommand{\latin}[1]{#1}
\makeatletter
\providecommand{\doi}
  {\begingroup\let\do\@makeother\dospecials
  \catcode`\{=1 \catcode`\}=2 \doi@aux}
\providecommand{\doi@aux}[1]{\endgroup\texttt{#1}}
\makeatother
\providecommand*\mcitethebibliography{\thebibliography}
\csname @ifundefined\endcsname{endmcitethebibliography}
  {\let\endmcitethebibliography\endthebibliography}{}

\end{document}